\documentclass[iop,apj]{emulateapj}
\newcommand{\Ms}{\ensuremath{M_{\odot}}}
\newcommand{\eg}{{\it e.g.}}
\newcommand{\cf}{{\it c.f. }}
\newcommand{\ie}{{\it i.e.}}
\newcommand{\viz}{{\it viz.}}

\slugcomment{In press with The Astrophysical Journal}

\shorttitle{Gas expulsion from infant R136 and NGC 3603}
\shortauthors{Banerjee and Kroupa}

\begin{document}

\title{Did the infant R136 and NGC 3603 clusters undergo residual gas expulsion?}

\author{Sambaran Banerjee and Pavel Kroupa}
\affil{Argelander-Institut f\"ur Astronomie, Auf dem H\"ugel 71, D-53121, Bonn, Germany;\\
sambaran@astro.uni-bonn.de, pavel@astro.uni-bonn.de}


\begin{abstract}
Based on kinematic data observed for very young, massive clusters that appear
to be in dynamical equilibrium, it has recently been argued that such young systems
set examples where the early residual gas-expulsion did not happen or had no dynamical effect. The intriguing
scenario of a star cluster forming through a single starburst has thereby been challenged.
Choosing the case of the R136 cluster of the Large Magellanic Cloud,
the most cited one in this context, we perform direct N-body computations that mimic the
early evolution of this cluster including the gas-removal phase (on a thermal timescale).
Our calculations show that under
plausible initial conditions as consistent from observational data,
a large fraction ($>60$\%) of a gas-expelled, expanding R136-like cluster is bound to
regain dynamical equilibrium by its current age. Therefore, the recent measurements
of velocity dispersion in the inner regions of R136, that indicate that the cluster
is in dynamical equilibrium, are consistent with an earlier substantial gas expulsion of R136
followed by a rapid re-virialization (in $\approx1$ Myr). Additionally, we find that the
less massive Galactic NGC 3603 Young Cluster (NYC), with a substantially
longer re-virialization time, is likely to be found deviated from dynamical equilibrium
at its present age ($\approx1$ Myr). The recently obtained stellar proper motions in the central part
of the NYC indeed suggest this and are consistent with the computed models. This work significantly extends
previous models of the Orion Nebula Cluster which already demonstrated that the re-virialization
time of young post-gas-expulsion clusters decreases with increasing pre-expulsion density.
\end{abstract}


\section{Introduction}\label{intro}

The question of how a young star cluster is formed has been debated for decades. According to the classical scenario,
a cluster forms essentially through a single star-burst event. The individual proto-stellar cores within the parent or proto-cluster gas cloud
approach their hydrogen-burning main-sequences (MS) to form an infant star cluster which still remains embedded within the residual gas that
did not collapse to stars.
This residual gas receives kinetic energy and radiation-pressure from the radiation of the massive MS and pre-main-sequence (PMS) stars
until it gets unbound from the system and then escapes.
This gas-removal process can be expected to be very rapid --- typically faster than or similar to the dynamical
crossing time of the embedded cluster \citep{lada2003}. The remaining gas-free cluster must expand due to the corresponding
dilution of gravitational potential well. For the hypothetical case of instantaneous gas removal,
the resultant cluster should get unbound if the total mass lost as gas is equal to or more
than the mass remaining in the stars (\ie, star formation efficiency (SFE) $\epsilon \leq 50$\%), as is true 
for any gravitationally self-bound system. For slower gas removal,
the survivability of the gas-deprived cluster as a bound system, for $\epsilon \leq 50$\%, increases \citep{pk2008}.
Two-body relaxation, which evolves the cluster towards a higher central concentration
until the beginning of gas-expulsion, also enhances survival.

The above scenario, applicable to the formation of Galactic and extra-galactic young star clusters, has been first depicted by, \eg,
\citet{hils80}, \citet{elm83} and \citet{mat83} either analytically or by small $N$ direct N-body studies \citep{lada84}.
These milestone works have later been elaborated by \citet{pketl2001}, \citet{gb2001}
and \citet{bk2007} by extensive direct N-body calculations of gas-expelling model infant clusters. Such theoretical studies,
along with observations of young embedded systems \citep{lada99,elm2000},
suggest a minimal star-formation efficiency of $\epsilon\approx 1/3$ for forming a bound, gas-free cluster,
applicable to embedded systems initially in dynamical (or virial) equilibrium.
An observational evidence of the residual gas expulsion process can be the $<2.5$ Myr old Orion Nebula Cluster (ONC)
whose measured velocity dispersion implies that it must be expanding \citep{jw88}. While the ONC was widely considered to end-up
in an unbound OB association (\eg, \citealt{zin93}), the detailed modelling by \citet{pketl2001} implies that
it is rather destined to become a bound cluster very similar to the Pleiades by 100 Myr.
Recent proper motion measurements of the Galactic $\approx 1$ Myr old NGC 3603
Young Cluster (NYC) by \citet{roch2010} also suggests that its stars might be away from energy-equipartition.
Notably, a demonstration of the expansion of clusters younger than a few Myr as they age is collated by \citet{brnd2008}.

The key properties of the stellar initial mass function (IMF; \citealt{pketl2012}), in particular, its observed universality
is also supported by a monolithic cluster formation scenario that involves competitive gas accretion by the most massive
proto-stellar cores. For example,
detailed three-dimensional, adaptive-mesh {\tt FLASH} calculations
including radiation feedback by  \citet{pet2010,pet2011} well reproduce not
only the general form of the universal IMF but also the relation between
the total mass (in stars) of an embedded cluster and the mass of its most massive stellar member \citep{wk2004}.
Recently, \citet{liu2012} find observational signatures of competitive accretion in the proto-cluster G10.6-0.4.
 
Very recently however, questions have been raised against the clustered mode of star formation, preferring
instead a hierarchical or continuous star formation (\eg, \citealt{brs2010,guter2011}). The latter is inferred from
the stellar density distribution in young, star forming regions over a ``global'' scale. Notably, \citet{pflz2012} showed that unless the
density contrast among the individual clusters is very high any such conclusions based on surface density profiles can be highly ambiguous and is
useful only when sufficiently deep and complete data are available. Furthermore, from the kinematics of several very young clusters
indicating their dynamical equilibrium, in particular that of the R136 cluster in the Large Magellanic Cloud (LMC), it has been argued that
they must have avoided any substantial residual gas-expulsion phase (\ie, effectively have had 100\% star-formation efficiency,
also see \citealt{gdwn2009} in this context); otherwise they
would have been found expanding at such a young age. This has been particularly emphasized very recently by
\citet{hb2012}.
A key ingredient that might be missing in such arguments is the consideration of the re-virialization of
the expanding gas-expelled cluster, \ie,
quick formation of a bound system after a fraction of the expanding cluster reverses back on a free-fall timescale.

The possible fate of the expanding ONC being a Pleiades-like bound system has been shown by \citet{pketl2001}
whose models already demonstrate that the re-virialization time decreases with increasing initial cluster density
(Fig.~1 of \citealt{pketl2001}). The survivability as a
bound cluster after gas-removal has been studied later in detail by, \eg, \citet{bk2007}. In this work, we find it crucial to study this
issue again with particular focus on the case of R136. This becomes essential because of the current
interpretations of the kinematic data of young, massive clusters such as R136, that seem to contradict the classical picture
of monolithic cluster formation and the associated event of residual gas expulsion.
We show that under plausible conditions,
the re-virialization of an R136-like massive cluster, after the gas-removal, is prompt enough to make it be found in virial equilibrium
despite of its young age. The recently obtained low (line-of-sight) velocity dispersion of single O-stars in R136 \citep{hb2012} is also
found to be consistent with a re-virialized system. We also find that the much lighter NYC, on the other hand,
is unlikely to be virialized at its current age of $\approx 1$ Myr \citep{stl2004,stl2006}, as indicated by \citet{roch2010}.

\section{Model computations}\label{compute}

\begin{table*}
\centering
\tablenum{1}
\caption{Computed model parameters. The ONC-A/B computations are from \citet{pketl2001} and included for comparison.
The quoted $\tau_{\rm vir}$s correspond to the re-virialization times of 30\% of the initial cluster mass (in stars) after
the delay time $\tau_d$.
The ``{\tt BSE}'' column indicates the presence of stellar evolution.}
\label{tab1}
\fontsize{10.0pt}{12.0pt}
\begin{tabular}{llllllllcc}
\\\hline
Cluster & $M_{\rm ecl}(0)/\Ms$ & $M_g(0)/\Ms$ & $r_h(0)/{\rm pc}$
 & $Z/Z_\odot$ & $\tau_g/{\rm Myr}$ & $\tau_{\rm cr}(0)/{\rm Myr}$ & $\tau_d/{\rm Myr}$
 & {\tt BSE} & $\tau_{\rm vir}$/Myr\\
\hline
R136 &  $1.0\times10^5$   & $2.0\times10^5$ &  $0.45$    & $0.5$         & $0.045$  & $0.021$   &  $0.0$, $0.6$    &  ${\rm Yes}$  & $0.9$\\
NYC  &  $1.3\times10^4$  &  $2.6\times10^4$ &  $0.34$    & $1.0$         & $0.034$  & $0.038$   &  $0.0$, $0.6$    &  ${\rm Yes}$  & $2.2$\\
\hline
ONC-A  & $3.7\times10^3$   &  $7.4\times10^3$ &  $0.45$    & $1.0$         & $0.045$  & $0.23$    &  $0.6$         &  ${\rm Yes}$  & $>10$\\ 
ONC-B  & $4.2\times10^3$   &  $8.4\times10^3$ &  $0.21$    & $1.0$         & $0.021$  & $0.066$   &  $0.6$         &  ${\rm Yes}$  & $\approx3$\\
\hline
\end{tabular}
\end{table*}

\subsection{Gas removal}\label{gasexp}

A thorough modelling of gas-removal from embedded clusters is complicated by the radiation hydrodynamical
processes which is extremely complex and involves uncertain physical mechanisms. For simplicity, we therefore mimic
the essential dynamical effects of the gas-expulsion process by applying a diluting, spherically-symmetric
external gravitational potential to a model cluster as in \citet{pketl2001}. Specifically, we use the potential
of the spherically-symmetric, time($t$)-varying mass distribution
\begin{eqnarray}
M_g(t)=& M_g(0) & t \leq \tau_d,\nonumber\\
M_g(t)=& M_g(0)\exp{\left(-\frac{(t-\tau_d)}{\tau_g}\right)} & t > \tau_d.
\label{eq:mdecay}
\end{eqnarray}
Here, $M_g(t)$ is the total mass in gas which is spatially distributed with the same initial Plummer density distribution
(Kroupa 2008; see below) as the stars and starts depleting with timescale $\tau_g$ after a delay of $\tau_d$.
The Plummer radius of the gas distribution is kept time-invariant \citep{pketl2001}.
Such an analytic approach is partially justified by \citet{gb2001} who perform comparison computations treating the gas
with the SPH method.

The exact values of the essential parameters quantifying the gas-expulsion timescale, \viz,
$\tau_g$ and $\tau_d$ depend on gas-physics. For simplicity, we use an average gas
velocity of $v_g\approx10$ km s$^{-1}$ which is the typical sound-speed in an
H II (ionized hydrogen) region. This gives
$$\tau_g=\frac{r_h(0)}{v_g},$$
where $r_h(0)$ is the initial half-mass radius of the stellar cluster/gas. The coupling of
stellar radiation with the ionized residual proto-cluster gas over-pressures the latter substantially and
can even make it radiation-pressure-dominated (RPD), for massive clusters in the initial
phase of the expansion of the gas. During the RPD phase, the gas is driven
out at speeds considerably exceeding the sound-speed of the ionized medium \citep{krm2009}. Once the expanding
gas becomes gas-pressure-dominated (GPD), it then continues to flow out
with the sound-speed of an H II region \citep{hils80}. Hence, the above $\tau_g$, from $v_g\approx10$ km s$^{-1}$,
represents its \emph{upper limit}; it can be shorter depending on the duration of the RPD state (also see Sec.~\ref{discuss}).
Notably, the initial RPD phase is crucial to launch the gas from very massive systems whose escape speed
exceeds the sound-speed in H II gas \citep{krm2009}.

As for the delay-time, we take the representative value of $\tau_d\approx0.6$ Myr
\citep{pketl2001}. The correct value of $\tau_d$ is again complicated by radiative gas-physics. An idea
of $\tau_d$ can be obtained from the lifetimes of the ultracompact H II (UCHII) regions which can be
upto $\approx 10^5$ yr (0.1 Myr; \citealt{chrch2002}). The very compact pre-gas-expulsion clusters (Sec.~\ref{dyn})
have sizes ($r_h(0)$) only a factor of $\approx3-4$ larger than the typical size of a UCHII region ($\approx0.1$ pc).
If one applies a similar Str\"omgren sphere expansion scenario (\citealt{chrch2002} and references therein)
to the compact embedded cluster, the estimated delay-time, $\tau_d$, before a sphere of radius $r_h(0)$
becomes ionized, would also be larger by a similar factor and hence close to the above representative value. Once the
gas is ionized, it couples strongly with the radiation from the O/B stars and launched immediately (see above).
High-velocity jet outflows from proto-stars \citep{pat2005} additionally facilitate the gas removal.

For super-massive clusters ($>10^6\Ms$), however, a ``stagnation radius'' can form
within the embedded cluster inside which the radiation cooling becomes sufficiently efficient to possibly
form second-generation stars \citep{wn2011}. Also, as discussed above,
the gas-outflow can initially be supersonic which generates shock-fronts. Although shocked,
it is unlikely that star-formation will occur in such an RPD gas. Later, during the GPD outflow, 
the flow can still be supersonic in the rarer/colder outer parts of the embedded cluster where the average
sound-speed might be lower than that typical for H II gas. However, it is not clear whether the cooling in the
shocked outer regions would be efficient enough to form stars.

Admittedly, the above arguments do not include complications such as unusual morphologies
of UCHIIs and possibly non-spherical ionization front, among others, and only provide basic estimates
of the gas-removal timescales. Observationally,
Galactic $\approx 1$ Myr old gas-free young clusters such as the ONC and the NYC imply that
the embedded phase is $\tau_d<1$ Myr. The above widely-used gas-expulsion model
does realize the essential dynamical effects
on the star cluster. In particular, such simplification has practically no effect on the remaining cluster after the gas
is expelled, \eg, on it's re-virialization, which is the focus of this work.

\subsection{Initial configuration, stellar dynamics and evolution}\label{dyn}

The initial model embedded stellar clusters are Plummer spheres \citep{pk2008} which the gas follows as
well. This is a reasonable approximation since dense interstellar medium (ISM)
filaments appear to have Plummer-like sections (\citealt{mali2012}; also see below).
The $r_h(0)$ of embedded clusters are substantially smaller than the typical sizes of exposed young clusters
as found in the semi-analytic calculations by \citet{mrk2012}.
These calculations constrain the birth-density of a large number of observed clusters by their
observed population of binary stars leading to a remarkable overlap with the densities of star-forming
molecular clumps (Fig.~6 in \citealt{mrk2012}).
Our initial clusters thus follow the empirical
relation between $r_h(0)$ and the embedded cluster mass (only the stars) $M_{\rm ecl}(0)$ by \citet{mrk2012}, \viz,
\begin{equation}
\frac{r_h(0)}{\rm pc}=0.10^{+0.07}_{-0.04}\times\left(\frac{M_{\rm ecl}(0)}{M_\odot}\right)^{0.13\pm0.04},
\label{eq:mrk}
\end{equation}
which is a rather weak dependence. This independently obtained result is in
excellent agreement with the observed results from \emph{Herschel}
(\citealt{andr2011}; Fig.~6.3). The star formation efficiency is taken to be $\epsilon=1/3$, \ie,
$M_g(0)=2M_{\rm ecl}(0)$ (see Sec.~\ref{discuss} for a discussion).

Notably, \citet{andr2011} and \citet{mali2012} refer to the shape and the compactness of ISM \emph{filament} sections. 
The star forming spherical/spheroidal proto-clusters form within the ISM filaments and at their intersections \citep{scn2012};
a part of the filament that collapses under self-gravity to form
an embedded cluster would become spherical/spheroidal. Recent \emph{Herschel} observations of ridges in molecular clouds
and of the associated filaments support this \citep{henne2012,scn2010,hil2011}.
The compact sizes ($\approx0.1$ pc) of these ISM filaments \citep{andr2011} then dictate the high compactness of the initial embedded systems.
There is, of course, the competing viewpoints whether all the stars form within one cloud (the standard scenario)
or the final cluster can be formed hierarchically (see Sec.~\ref{intro}).
Currently, the possibility of forming massive stars within the filaments but outside any cluster
has also been suggested (\eg, \citealt{brs2012}). So far our computations in this paper are concerned,
we adopt the standard scenario and investigate whether the observed kinematics of the young clusters like R136 and NYC (to start with)
can be reasonably explained within such a context.
In that case, our chosen initial conditions well reflect the sizes and the shapes of the observed filaments.

The IMFs of the clusters are chosen to be canonical \citep{pk2001} with the most massive star following the $m_{\rm max}-M_{\rm ecl}(0)$ relation
of \citet{wk2004}. To that end an ``optimized sampling'' algorithm \citep{pketl2012} is used instead of randomly sampling the
IMF. All the computed models are fully mass-segregated using the method of \citet{bg2008}, as observed in young clusters
\citep{lit2003,chen2007,pz2010}. For computational ease, we take all the members of the initial cluster to be single stars. Since
the O-stars in young clusters are typically found in binaries \citep{sev2011,sana2012},
our initial stellar population, admittedly, does not completely represent
that of a young cluster. However, binarity is {\it unlikely} to substantially influence the expansion of a cluster during gas-expulsion
and re-collapse and virialization thereafter, which is the focus of this work. 
Although the initial segregated state is detected for several Galactic young clusters, it is
yet to be confirmed for young clusters in general (see \citealt{pz2010} for a discussion).
Therefore, for comparison purposes, we compute identical models
\emph{without} any initial mass-segregation. We
find that mass-segregation does not influence the conclusions (Sec.~\ref{r136}).

The dynamical evolution of the model clusters are computed using the state-of-the-art {\tt NBODY6} code \citep{ar2003}.
In addition to integrating the particle
orbits using the highly accurate fourth-order Hermite scheme and dealing with the diverging gravitational
forces in close encounters through regularizations, {\tt NBODY6} also employs the well-tested analytical stellar and binary
evolution recipes of \citet{hur2000,hur2002}, \viz, the {\tt SSE} and the {\tt BSE} schemes.
{\tt NBODY6} also includes the time-variable Plummer gas potential as described above. 

The initial masses of the computed model embedded clusters are chosen to be $M_{\rm ecl}(0)\approx10^5\Ms$ which is an upper mass limit
of R136 \citep{crw2010}. Their Plummer radii are chosen from Eqn.~(\ref{eq:mrk}). We also consider
an embedded cluster of $M_{\rm ecl}(0)=1.3\times10^4\Ms$ to mimic the gas-removal from a Galactic NYC-like cluster \citep{stl2004,stl2006,roch2010}.
The metallicities of the former models are taken to be $Z=0.5Z_\odot$, as appropriate for the LMC, and that for the NYC model it is $Z=Z_\odot$.
Our computed models are summarized in Table~\ref{tab1}, where those in \citet{pketl2001} are also added for comparison. For
both of our computed models, the $\tau_g$s are similar to the initial crossing times $\tau_{\rm cr}(0)$;
$\tau_g\approx$ 2 and 1 time(s) $\tau_{\rm cr}(0)$ for the R136 and the NYC model respectively. Since $\tau_g<\tau_{\rm cr}(0)$
for the \citet{pketl2001} models, our gas-expulsions are less ``prompt'' owing to higher initial concentrations.
The {\tt MCLUSTER} program \citep{kup2011} is used to set the initial configurations.
As it is difficult to model the weak tidal field of the LMC, the clusters are evolved in absence of any galactic potential.

\section{Results}\label{res}

\subsection{Gas-expulsion from an R136-like cluster}\label{r136}

\begin{figure}
\centering
\includegraphics[height=6.5cm,angle=0]{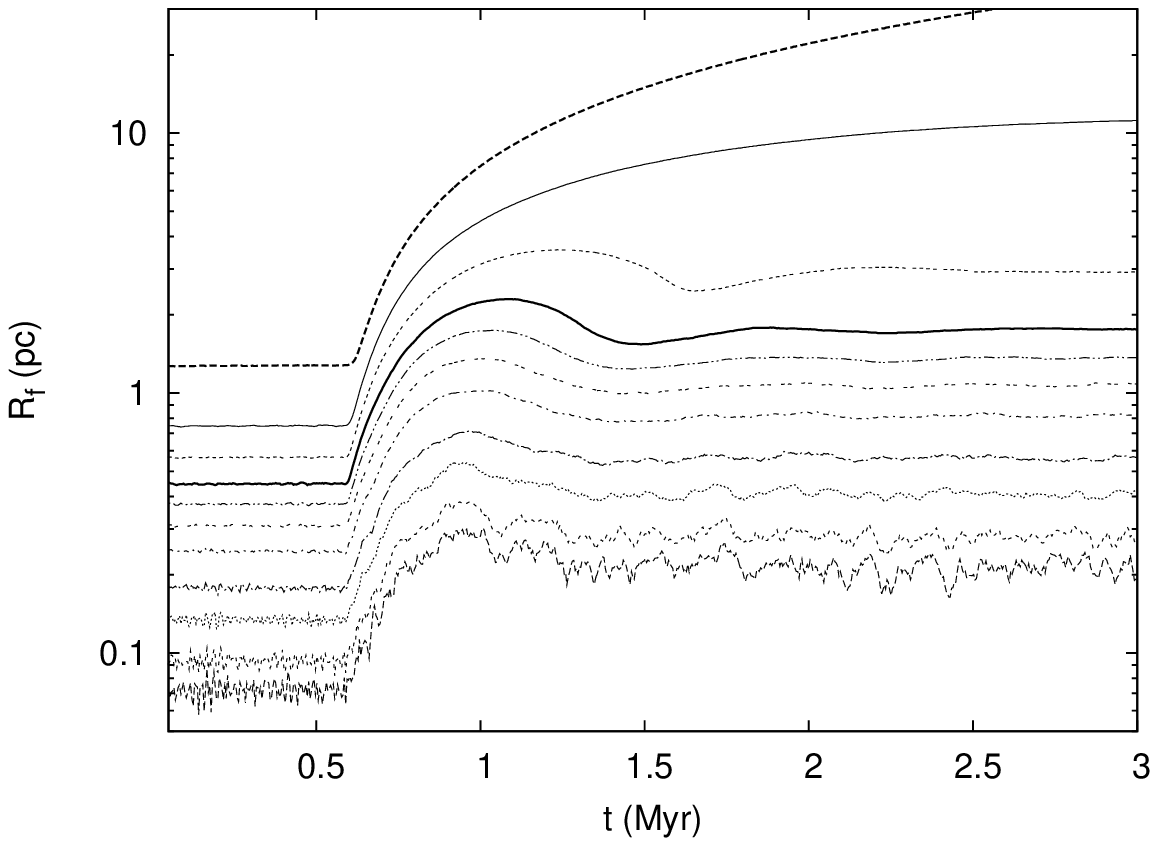}
\includegraphics[height=6.5cm,angle=0]{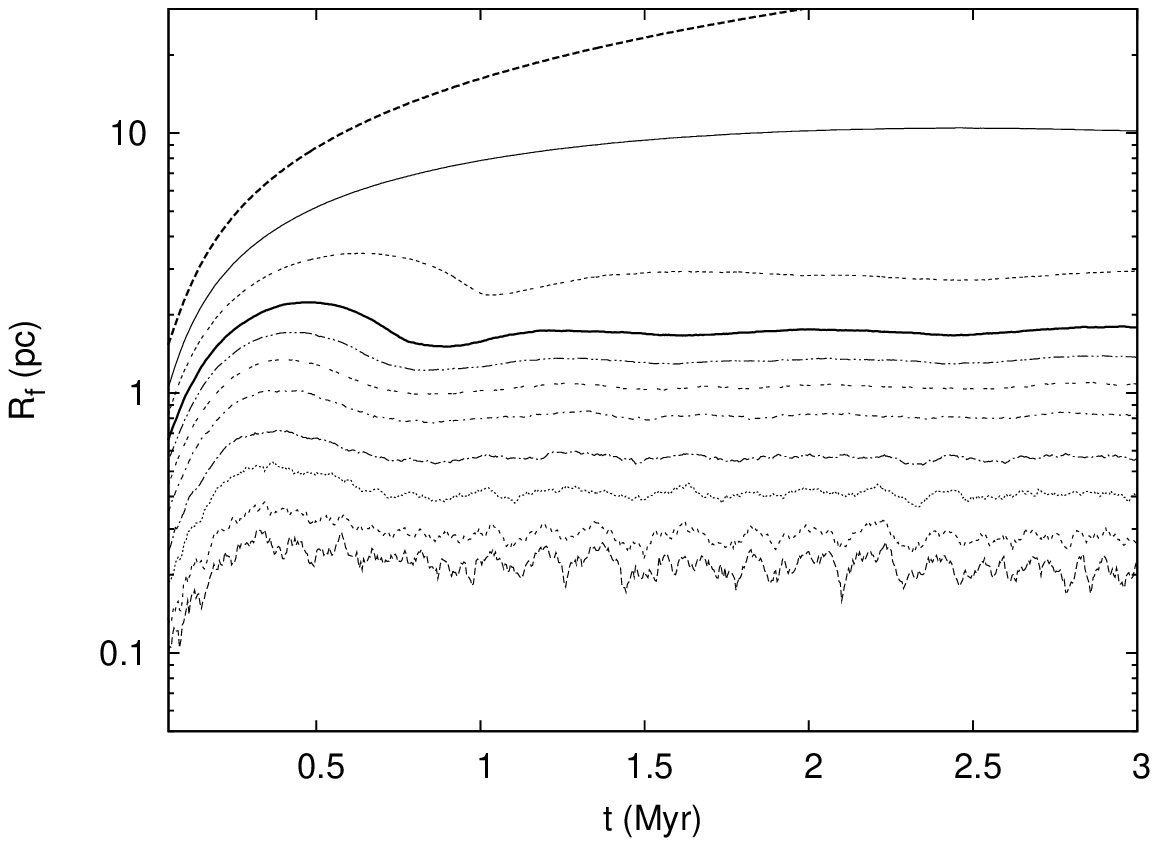}
\caption{The evolution of the Lagrange radii, $R_f$, for stellar cluster mass fractions $f$
(escaping stars inclusive) for the computed R136 models
with $\tau_d=0.6$ Myr (top) and $\tau_d=0$ (bottom). In each panel, the curves, from bottom to top,
correspond to $f=$ 0.01, 0.02, 0.05, 0.1, 0.2, 0.3, 0.4, 0.5, 0.625, 0.7 and 0.9 respectively. The thick solid line
is therefore the half mass radius of the cluster.}
\label{lrad_R136}
\end{figure}

\begin{figure}
\centering
\includegraphics[height=6.5cm,angle=0]{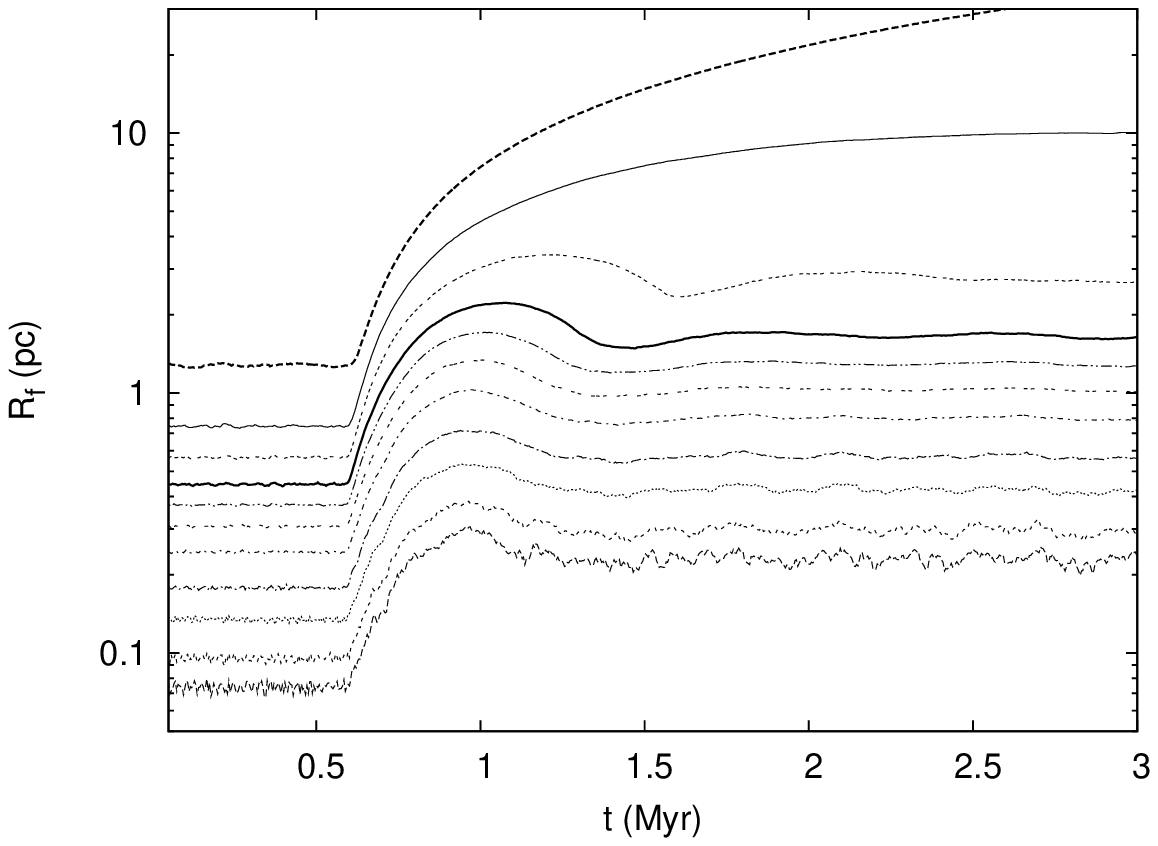}
\includegraphics[height=6.5cm,angle=0]{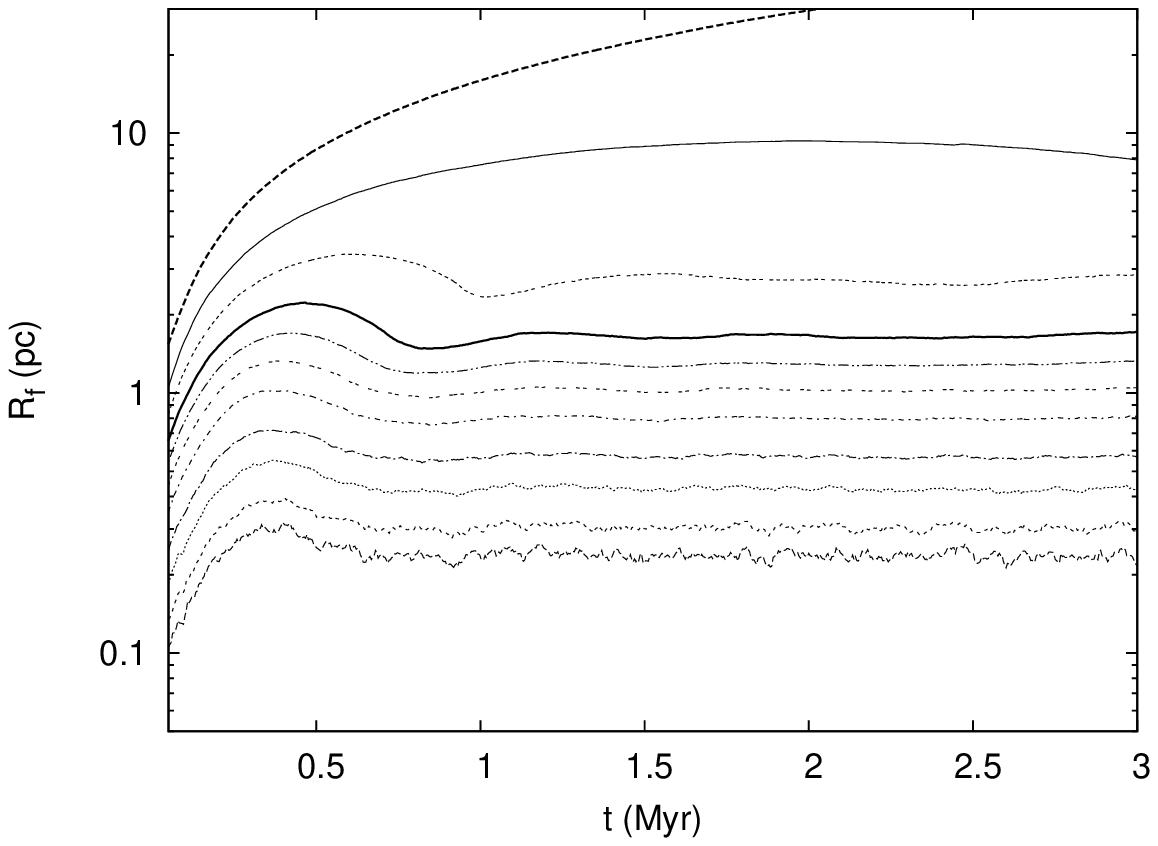}
\caption{The evolution of Lagrange radii, $R_f$, for computed models without any initial mass-segregation but
otherwise identical to the R136 models in Fig.~\ref{lrad_R136}. The legends are the same as in Fig.~\ref{lrad_R136}.}
\label{lrad_R136_noseg}
\end{figure}

\begin{figure}
\centering
\includegraphics[height=6.5cm,angle=0]{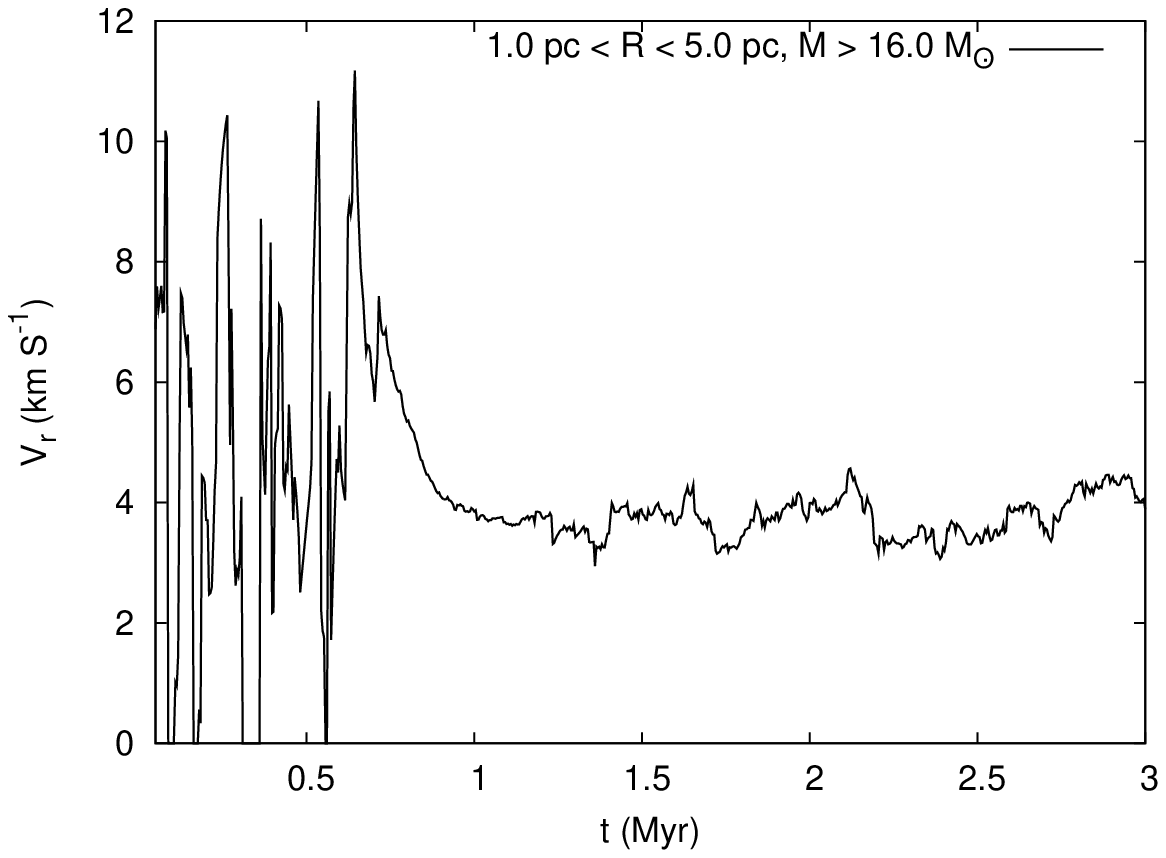}
\includegraphics[height=6.5cm,angle=0]{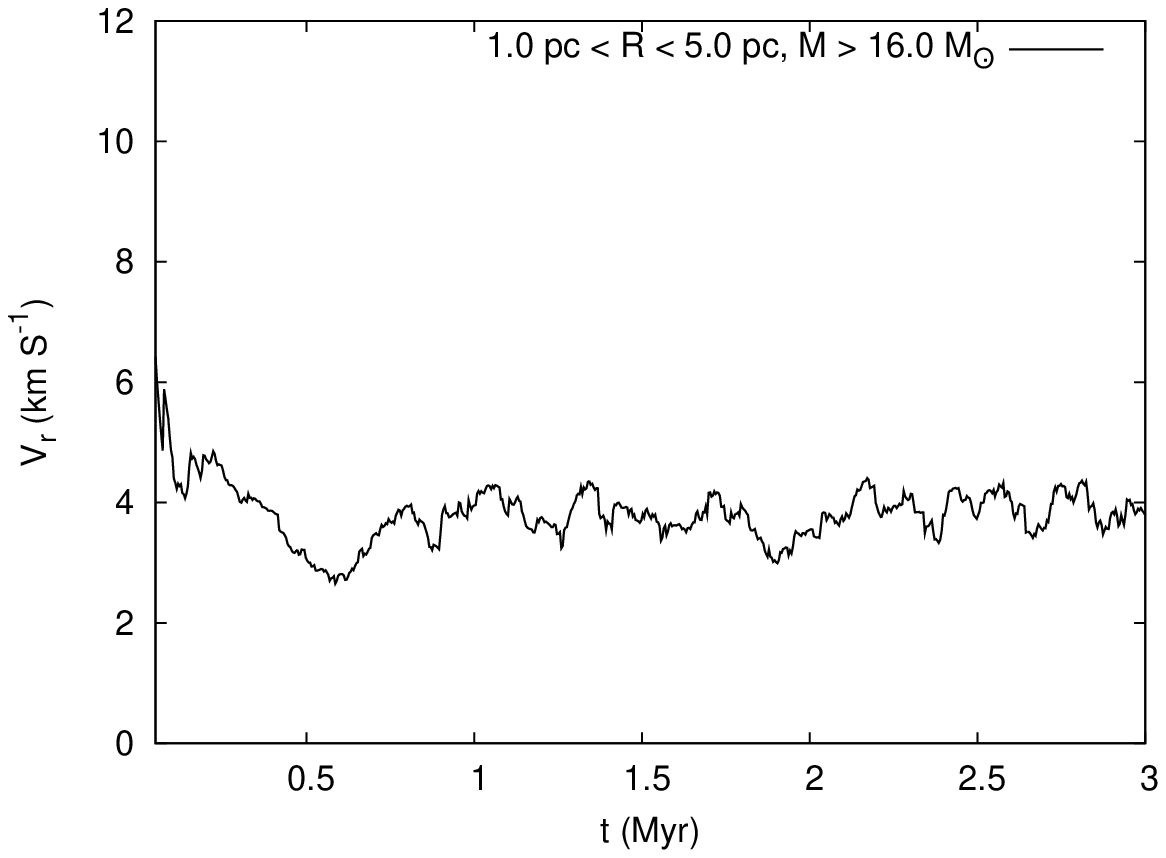}
\caption{The evolution of the radial velocity (RV) dispersion, $V_r$, of the O-stars ($M>16\Ms$), within the projected distances
$1{\rm~pc}< R<5{\rm~pc}$ from the cluster center, for the computed R136 models. The top and the bottom panels correspond to the
computations with $\tau_d=0.6$ and 0 Myr respectively.}
\label{vdis_R136}
\end{figure}

Fig.~\ref{lrad_R136} shows the evolution of the Lagrange radii, $R_f$, in our computed R136 model clusters
(Table~\ref{tab1}) with $\tau_d=0.6$ (top panel) and 0 (bottom)
Myr. The $\tau_d=0$ case is computed as a test case only to compare with that of the more realistic delay $\tau_d=0.6$ Myr.
From the beginning of the gas-expulsion, the cluster expands rapidly with timescale $\tau_g$. At least 60\% of the cluster, by mass,
then collapses back to a steady size, \ie, regains virialization in $\tau_{\rm vir}\approx1$ Myr (beyond $\tau_d$).
The inner regions virialize even earlier. Given that the bulk of the R136
cluster is $\approx 3$ Myr old \citep{and2009}, it ought to be currently in dynamical equilibrium as a whole \emph{even if
it has undergone a substantial gas-expulsion phase in the past}. We discuss this further in Sec.~\ref{discuss}.
The gas-free cluster is expanded
by a factor of $\approx 3$ in terms of its half-mass radius, $r_h$, after reverting to dynamical equilibrium. Notably, \emph{the re-virialization
happened in spite of the wind mass-loss from the massive stars} that was operative during these computations. Comparison
of the two panels in Fig.~\ref{lrad_R136} makes it apparent that a finite $\tau_d$ (with $\tau_g$ unchanged) keeps
the form of the cluster's expansion and re-virialization identical --- it merely applies a time-translation to the overall
evolution.

Fig.~\ref{lrad_R136_noseg} shows the $R_f$-evolution for identical R136 models but without initial mass-segregation.
It demonstrates that the cluster
evolutions are throughout identical to the previous cases with primordial mass segregation, in particular during the expanding phase
and the re-collapse to virialization. In other words, primordial mass-segregation has no effect on the re-virialization of a cluster.

Very recently, \citet{hb2012} measured radial/line-of-sight velocities (RV) of \emph{single} O-stars within
$1{\rm~pc}\lesssim R\lesssim5{\rm~pc}$ projected distance from R136's center\footnote{Strictly, the distances were
measured from the most massive cluster member star R136a1. The exact location of the R136's true density center being unknown,
we consider this star be expectedly very close to the cluster's density center.}
 with data obtained from the
``VLT-FLAMES Tarantula Survey'' (VFTS; \citealt{evn2011}). They conclude that the RV dispersion, $V_r$, of the single O-stars within this
region is $4{\rm ~km~s}^{-1}\lesssim V_r\lesssim5{\rm ~km~s}^{-1}$, conforming with R136 being in virial equilibrium.

Fig.~\ref{vdis_R136} shows the evolution of $V_r$ corresponding to the R136 computations in Fig.~\ref{lrad_R136}
for $1{\rm~pc}< R<5{\rm~pc}$ projection from the density center and stellar masses $M>16\Ms$,
\ie, those of the O-stars. It demonstrates that, after re-virialization,
the O-stars indeed have $V_r$ remarkably similar to that measured by \citet{hb2012}. The initial large fluctuations in $V_r$ in the upper
panel of Fig.~\ref{vdis_R136} ($\tau_d=0.6$ Myr) is due to the initial mass-segregated condition which results in only a few O-stars  
within $1{\rm~pc}< R<5{\rm~pc}$ initially. The measured RV data is therefore consistent with R136 being through a gas-expulsion phase
and then in a re-virialized state now, \emph{unlike} claimed by \citet{hb2012}.

\subsection{The case of the NGC 3603 Young Cluster}\label{nyc}

\begin{figure}
\centering
\includegraphics[height=6.5cm,angle=0]{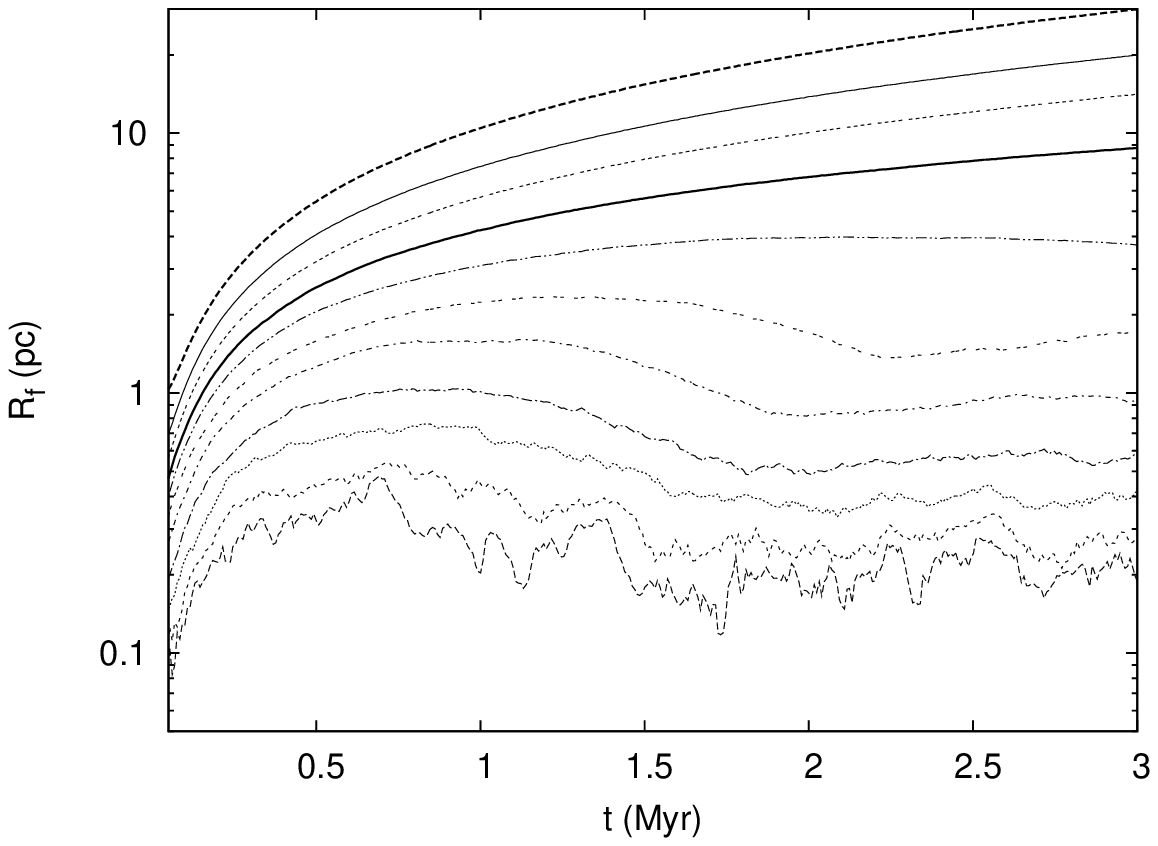}
\includegraphics[height=6.5cm,angle=0]{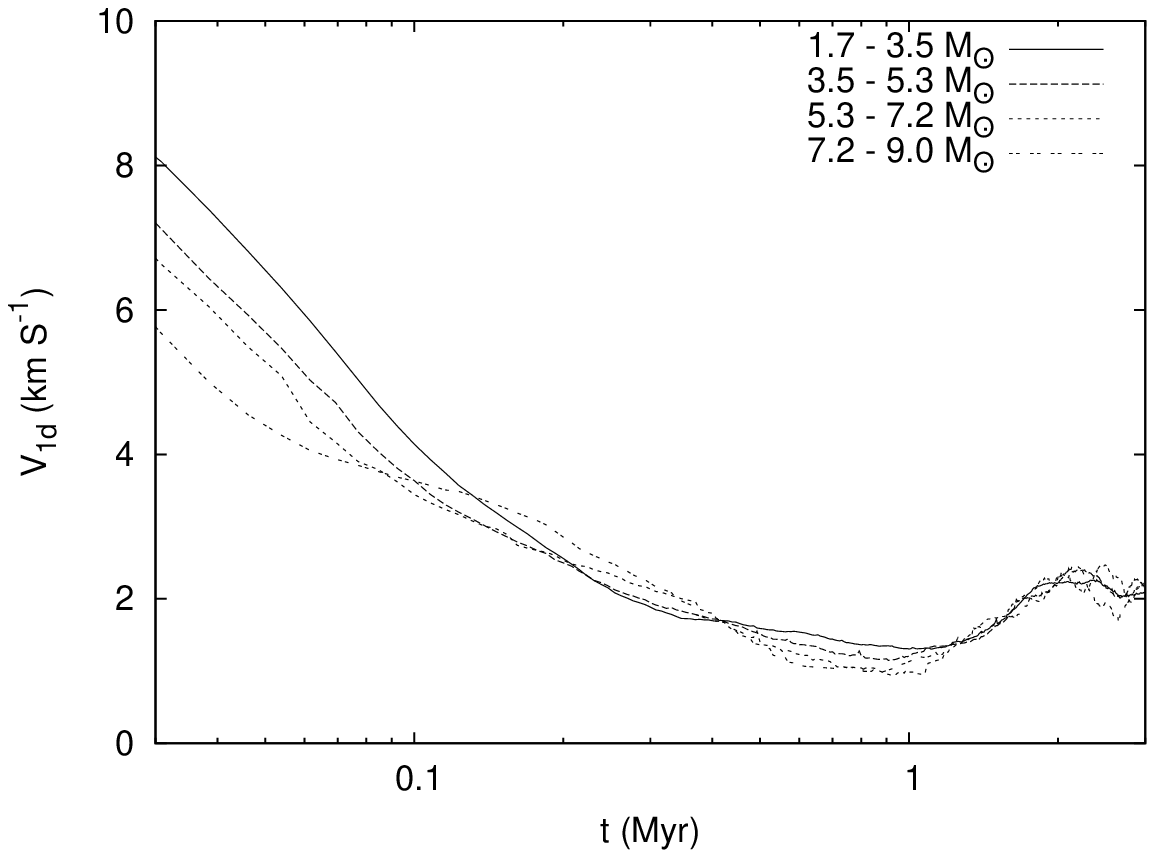}
\caption{Top: The evolution of the Lagrange radii for the computed NGC 3603 Young Cluster model with $\tau_d=0$. The legends are the same as in
Fig.~\ref{lrad_R136}. Bottom: The variation of the 1-dimensional (transverse) velocity dispersion, $V_{\rm 1d}$, within $R<1$ pc projection,
for 4 mass-bins chosen within the same mass range as in \citet{roch2010}, for the above model. Note that initially $V_{\rm 1d}$ differs
between the mass bins consistent with energy equipartition. This separation of $V_{\rm 1d}$ values vanishes as a result of violent relaxation
driven by gas-expulsion. Note that the $t$-axis of the bottom panel is plotted in the logarithmic scale to highlight this.}
\label{lrad_NYC}
\end{figure}

\begin{figure}
\centering
\includegraphics[height=6.5cm,angle=0]{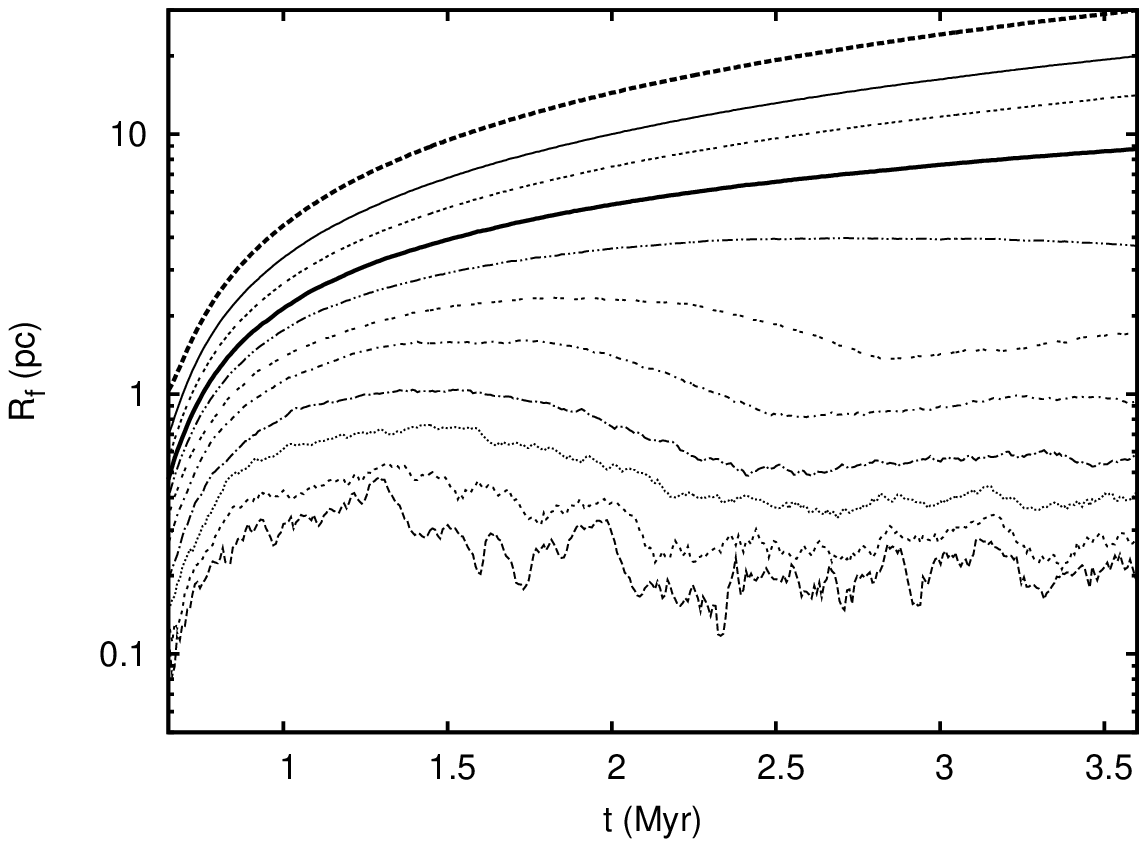}
\includegraphics[height=6.5cm,angle=0]{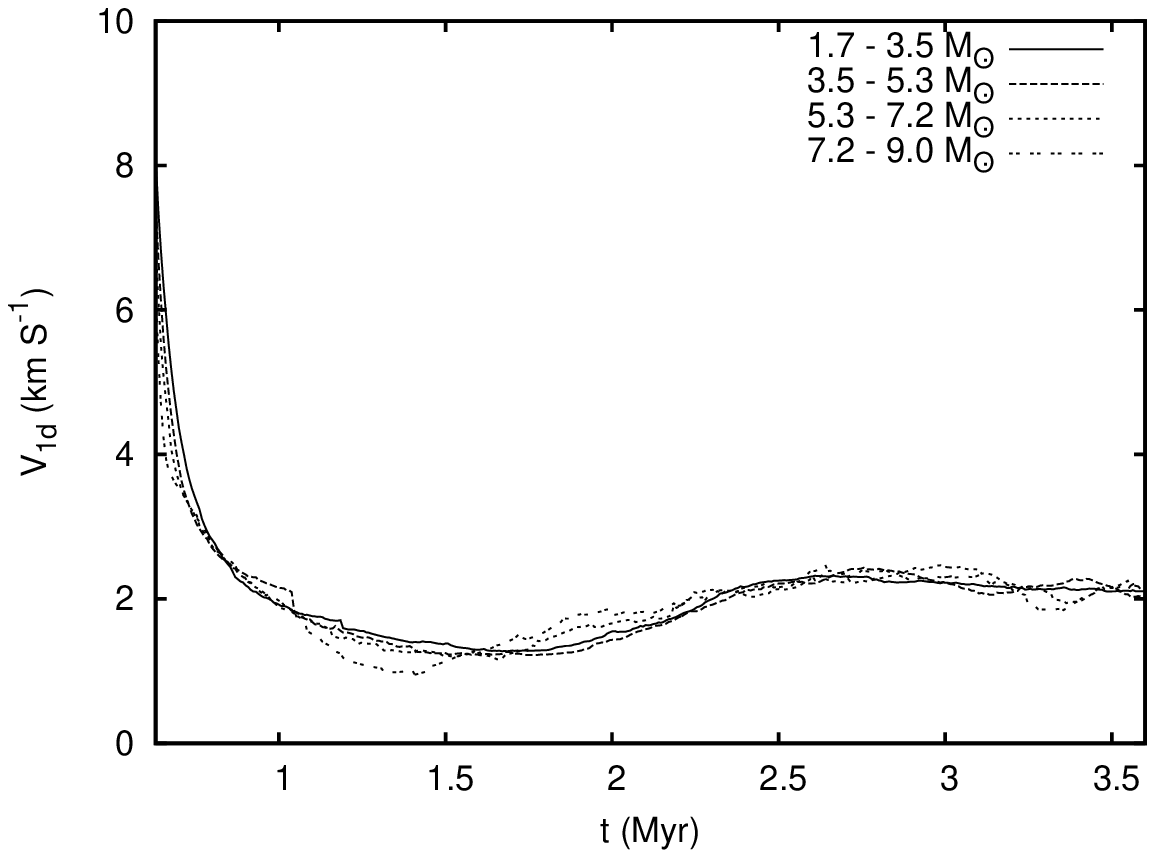}
\caption{Evolutions of the Lagrange radii (top) and 1-dimensional velocity dispersion (bottom) of the same computed NYC models as in
Fig.~\ref{lrad_NYC} where an initial gas expulsion time-delay of $\tau_d=0.6$ Myr is applied while plotting. Hence, the
plots are essentially the same as in Fig.~\ref{lrad_NYC} except that they begin at $t=0.6$ Myr and correspond to those
from a computation with $\tau_d=0.6$ Myr (see Sec.~\ref{r136}). They show that NYC
would be super-virial at its current $\approx 1$ Myr age, as observed.}
\label{lrad_NYC2}
\end{figure}

Fig.~\ref{lrad_NYC} (top) shows the Lagrangian radii of our computed NYC cluster (Table~\ref{tab1}). The mass of NYC
\citep{stl2004,stl2006,roch2010} is substantially smaller than that of R136, implying a longer free-fall and hence
re-virialization time. The model is computed with $\tau_d=0$, which is rather unrealistic, to see the effect of
the earliest possible gas removal. Even then the cluster is still expanding in its outer parts and just beginning
to fall back in the inner regions ($\lesssim R_{0.3}$) at its very young age of $\approx 1$ Myr \citep{stl2004}, as
can be seen in Fig.~\ref{lrad_NYC}. In other words, it is practically impossible for NYC to be in
dynamical equilibrium at its present age had it undergone a substantial gas-expulsion phase earlier. The computed cluster
re-virializes in $\tau_{\rm vir}>2$ Myr ($<40$\% of it; \cf Fig.~\ref{lrad_NYC})
as opposed to our R136 models (Sec.~\ref{r136}) which take
$\approx1$ Myr for the same ($>60$\% of them; \cf Fig.~\ref{lrad_R136}, Table~\ref{tab1}).

From proper motion measurements within the central $\approx1$ pc projection of NYC, \citet{roch2010} found no notable differences
among the 1-dimensional velocity dispersions ($V_{\rm 1d}$) of 4 mass (magnitude) bins over $1.7-9\Ms$. This indicates that NYC is far from
energy equipartition as it should be if the system is out of dynamical equilibrium (the converse conclusion that non-equipartition
implies non-equilibrium, as used by \citealt{roch2010},
is however \emph{not} necessarily true). This is explicitly demonstrated in Fig.~\ref{lrad_NYC} (bottom),
where the initial separation in $V_{\rm 1d}$ is practically vanished at $t=1$ Myr due to violent relaxation driven
by the gas-expulsion.

The computed $V_{\rm 1d}$ at $t=1$ Myr is somewhat
less than the measured\footnote{The $V_{\rm 1d}$ is obtained from
proper motion measurements and is therefore ``binary-corrected''.}
$V_{\rm 1d}\approx4.5\pm0.8{\rm~km~s}^{-1}$ and the observed system must currently be super-virial,
unlike the computed sub-virial state at $t=1$ Myr, as the inferred dynamical mass\footnote{The dynamical
$1.7\times10^4\Ms$, doesn't represent an upper limit as quoted by \citet{roch2010}, but is only an individual estimate.}
exceeds the photometric mass \citep{roch2010}. These can be easily accounted for by
a plausible $0.6<\tau_d<0.8$ Myr delay in the gas removal (\cf Fig.~\ref{lrad_NYC})
which would then result in agreement with the observed $V_{\rm 1d}$ and the super-virial
state of NYC. Furthermore, NYC's mass is uncertain by a factor $\approx 2$ \citep{roch2010}.

Fig.~\ref{lrad_NYC2} replots Fig.~\ref{lrad_NYC} with a $\tau_d\approx0.6$ Myr delay which, in turn, corresponds to the
results of a computation with this delay in gas-expulsion (see Sec.~\ref{r136}). This makes the cluster
super-virial at its $\approx 1$ Myr current age. The corresponding $V_{\rm 1d}$ is still somewhat smaller than the observed value
but is within $3\sigma$ limits. Notably, NYC might, in fact, be somewhat younger (\cf Fig.~4 of \citealt{stl2004}) and
$V_{\rm 1d}$ also depends on the exact initial mass and size which are subject to uncertainties. Moreover, NYC's $V_{\rm 1d}$
and age are susceptible to its distance uncertainties.
Notably, \citet{cott2012} also find the Westerlund I cluster sub-virial.

\section{Conclusions and outlook}\label{discuss}

The key message from our above computations and analyses is a reminder that \emph{an observed dynamical equilibrium state of a
very young stellar cluster does \emph{not} necessarily dictate that the cluster has \emph{not} undergone a gas-expulsion phase}. The non-occurrence
of gas-removal has been highlighted, particularly by putting forward the case of R136, by recent authors, \eg, \citet{hb2012}.
This is why we focus on R136 in this study. We also demonstrate, choosing the example of NGC 3606, that
lower-mass clusters take much longer to re-virialize (\cf Table~\ref{tab1}) so that it is as well appropriate to find non-equilibrium
signatures in them. This has been shown to be true also for the ONC by \citet{pketl2001} (\cf Table~\ref{tab1}). The above must be
remembered while interpreting the kinematics of any young cluster.

Notably, R136 exhibits an age-spread; its massive stellar population can be younger, \viz, $\lesssim2$ Myr old \citep{mhunt98}.
If the gas-expulsion is assumed to commence
only after the formation of the most massive stars, a currently equilibrium state is still plausible given the $\tau_{\rm vir}\approx1$ Myr
re-virialization time. It should, however, be remembered that age-estimates of massive stars are highly uncertain. There
isn't any concrete evidence that invalidates the scenario involving the formation of R136's entire stellar population
at once $\approx3$ Myr \citep{and2009} ago and the most massive single stars forming later via dynamically induced
binary mergers \citep{bko2012b} or appearing younger due to mass transfer in close O-star binaries.

Although the present study does not cover a general range of the parameters $\tau_d$, $\tau_g$, $r_h(0)$ and
$\epsilon$, these computations can still provide a fair idea of the effects of the variation of these parameters over their
plausible ranges. By applying time-shifts to the point of expansion of the clusters in Figs.~\ref{lrad_R136} \& \ref{lrad_R136_noseg},
it can be concluded that $\tau_d\lesssim2$ Myr required to find R136 virialized at $t=3$ Myr, since $\tau_{\rm vir}\approx1$ Myr.
Although an unambiguous $\tau_d$ requires radiation hydrodynamic calculations, this upper limit is perhaps too large a value
for $\tau_d$ in the light of the discussions in Sec.~\ref{gasexp}. The currently chosen values of $\tau_g$s are upper limits based on
the typical H II-gas sound-speed of $\approx10$ km s$^{-1}$ (Sec.~\ref{gasexp}). As discussed in Sec.~\ref{gasexp},
the outflow speed can initially be significantly higher than the sound-speed, driven mainly by radiation pressure,
resulting in smaller $\tau_g$s than those assumed. This would make the gas-expulsion more prompt,
keeping $\tau_{\rm vir}$ practically unchanged.

As for the SFE, the chosen $\epsilon\approx1/3$ is in the range of the observationally estimated SFEs by
\citet{lada2003} (see their Table~2). Recent theoretical modelling of star formation with high-resolution
resistive magnetohydrodynamics \citep{mact2012} also suggest $\epsilon\approx33\%$. 
In any case, a larger $\epsilon$ would also result in a bound system and shorter
$\tau_{\rm vir}$ as the re-virializing mass would be larger. As for $r_h(0)$s, their increasing values would increase
$\tau_{\rm vir}$s, so our conclusions \emph{depend} on our adopted initial high compactness. However, this condition
is plausible since such compact star-forming environments
(a factor of $\approx10$ smaller than typical young clusters, \ie, $r_h(0)\approx0.3-0.4$ pc)
are inferred observationally by \citet{andr2011} and as well from the semi-analytic study by \citet{mrk2012} independently
(the re-virialization time depends on the above typical compactness but, of course, not on the specific Eqn.~(\ref{eq:mrk})).
In other words, plausible variations of the model parameters are unlikely to alter our above primary conclusions since none
imply a substantial lengthening of $\tau_d$, $\tau_g$ or $\tau_{\rm vir}$.

An immediate improvement over the present study would be to introduce an appropriate binary population as in
\citet{bko2012a,bko2012b}. In the light of the present and upcoming observations of Galactic and extragalactic young
clusters, an important pending task is to systematically and quantitatively
study the effect of gas-expulsion over their entire mass-range
with varying parameters such as $\tau_d$, $\tau_g$, $r_h(0)$ and $\epsilon$ including realistic stellar populations. Such a model-bank
would be extremely valuable to interpret the ever-enriching kinematic data of young stellar clusters.

\acknowledgements

We thank the anonymous referee for useful and guiding comments which helped in improving the manuscript.  
We are thankful to Jan Pflamm-Altenburg of the Argelander-Institut f\"ur Astronomie, Bonn, Germany and Carsten Weidner
of the Instituto de Astrof\'isica de Canarias, La Laguna, Spain for motivating discussions.

\end{document}